\documentclass[preprint,12pt]{elsarticle}

\usepackage{amssymb}
\usepackage{graphicx}
\usepackage{subfigure}

\begin{document}

\begin{frontmatter}



\title{Simulation on the Transparency of Electrons and Ion Back Flow for a Time Projection Chamber based on Staggered Multiple THGEMs}


\author[a]{Mengzhi Wu}
\ead{wumengzhi17@mails.ucas.edu.cn}
\author[a]{Qian Liu}
\ead{liuqian@ucas.ac.cn}
\author[a]{Ping Li}
\author[a]{Shi Chen}
\author[a]{Binlong Wang}
\author[a]{Wenhan Shen}
\author[a]{Shiping Chen}
\author[a]{Yangheng Zheng}
\author[b]{Hongbang Liu}
\author[c]{Yigang Xie}
\author[c]{Jin Li}

\affiliation[a]{organization={School of Physics, University of Chinese Academy of Sciences},
            addressline={}, 
            city={Beijing},
            postcode={100049}, 
            state={},
            country={China}}

\affiliation[b]{organization={School of Physics, Guangxi University},
            addressline={}, 
            city={Nanning, Guangxi},
            postcode={530004}, 
            state={},
            country={China}}

\affiliation[c]{organization={Institute of High Energy Physics},
             addressline={},
             city={Beijing},
             postcode={100049},
             state={},
             country={China}}

\begin{abstract}
The IBF and the transparent rate of electrons are two essential indicators of TPC, which affect the energy resolution and counting rate respectively. In this paper, we propose several novel strategies of staggered multi-THGEM to suppress IBF, where the geometry of the first layer THGEM will be optimized to increase the electron transparent rate. By Garfield++ simulation, the electron transparency rate can be more than 90\% of single THGEM with a optimized large hole. By simulating these configurations of triple and quadruple THGEM structures, we conclude that the IBF can be reduced to 0.2\% level in an optimized configuration denoted as "ACBA". This strategy for staggered THGEM could have potential applications in future TPC projects.
\end{abstract}



\begin{keyword}


staggered multiple THGEM, ion back flow, electron transparent rate, Garfield++
\end{keyword}

\end{frontmatter}


\section{Introduction}
The time projection chamber (TPC) is an important detector in particle physics and nuclear physics experiments. Gaseous TPCs are widely used in nuclear collider experiments like STAR\cite{bib: STAR} and ALICE\cite{bib: ALICE, bib: ALICE-Upgrade}, and electron collider experiments like ALEPH\cite{bib: ALEPH} retired and CEPC\cite{bib: CEPC} in progress, while liquid or dual-phase TPCs are used in neutrino experiments NEXT\cite{bib: NEXT}, and dark matter experiments XENON-1T\cite{bib: XENON-1T} and PANDAX\cite{bib: PandaX}. The basic function of TPC is to measure the 3-dimensional track of final-state particles. In a collider experiment, TPC is able to measure charge and momenta of particles by a deflection magnetic field. In addition, for a GeV scale experiment, TPC can also be used for particle identification (PID) by measuring the energy loss dE/dx of particles in working gas.
	
Modern gaseous TPC usually use micro pattern gaseous detectors (MPGD)\cite{bib: MPGD} as their readout chamber, in order to adapt the high counting rate $\rm MHz/mm^2$ in modern colliders. The main technical indicators of a MPGD contains the gain, the transparency of electrons, the ion back flow (IBF) rate and the counting rate. These factors would influence its performance such as the detection efficiency, the spacial resolution and the energy resolution. 

The transparency of electrons is the ratio at which the electrons produced by the primary ionization can pass through the MPGD. This ratio affects the energy resolution of TPC, so it impacts the measurements on the energy loss dE/dx of particles, which is essential for particle identification(PID) of Pions, Kaons and protons with momenta less than 1\,GeV. Hence the transparent rate of electrons is expected to be as large as possible. In reference \cite{bib: GEM-transparency}, the authors studied several GEM-like structures by Garfield simulation and optimized the electron transparency to more than 80\%. 

The ions produced by primary and secondary ionization would flow back to the drift zone of a TPC. These ions will lead the spacial charge effect that the drift electric field is distorted which would affect the velocity of drift electrons and result in the bad spacial resolution of TPC. Therefore, the IBF in a TPC is expected to be suppressed as low as possible. One solution is to seek new structures of MPGD. For example, the mesh of MicroMegas can absorb a number of ions, so some groups try to use double and triple meshes structures to reduce ions with IBF 0.01\% level\cite{bib: DMM-IBF-1,bib: DMM-IBF-2}. However, it is hard for MicroMegas to achieve large-size manufacturing. For example, the size MicroMegas in \cite{bib: DMM-IBF-1,bib: DMM-IBF-2} is $\rm 2.5\,cm\times 2.5\,cm$, which is too small for a large experiment.

THick GEMs (THGEM), introduced in parallel by different groups \cite{bib: THGEM-1, bib: THGEM-2} are electron multipliers derived from the GEM design, scaling the geometrical parameters and changing the production technology. Typical values of the geometrical parameters are kapton thickness of 0.4-1.0\,mm, hole diameter ranging between 0.3 and 1.0\,mm, hole pitch of 0.7-1.2\,mm and rim width between 0 and 0.1\,mm. The working electric fields usually contains the drift field, the avalanche field in the hole, the transfer field between two THGEM, and the induction field between bottom THGEM and the anode. THGEM has some advantages like the large gain, the robustness, the large size, and mechanical characteristics. Recently, several groups \cite{bib: Staggered-THGEM-UCAS, bib: Staggered-THGEM-SDU} studied the staggered multi-THGEM configurations to suppress the IBF, and the latest result is 0.58\% to 0.71\% with a stable gain $\sim$2500 \cite{bib: Staggered-THGEM-SDU}.

In this work, we propose THGEM with large hole diameter to improve the transparency rate of electrons and multi-THGEM with novel staggered strategies to suppress the IBF. In section 2, we simulate the drift velocity and the transverse diffusion coefficients of electrons in different gases to optimize the gas ratio and the electric field in drift area. In section 3, we simulate the transparency of electrons of a single THGEM, and study the influence of the hole size of THGEM and the electric field, and finally determine the diameter as 0.6\,mm where the transparency can achieve 90\% in Ar-$\rm iC_4H_{10}$(90/10). In section 4, we propose several staggered configurations of triple and quadruple THGEMs and find out that the IBF can be reduced to 0.01\% level in a optimized configuration.

\section{The gas properties}

It is important to choose an appropriate working gas to optimize the performance of TPC. In laboratory, the binary gas mixtures are commonly used such as Ar-$\rm CF_4$ \cite{bib: Ar-CF4}, Ar-$\rm CO_2$ \cite{bib: Ar-CO2} Ar-$\rm iC_4H_{10}$ \cite{bib: DMM-3} as well as some ternary mixtures. The drift velocity and diffusion of electrons are quite different in pure gases. Generally, the electron drift velocity is supposed to be less affected by electric field in order to optimize the time and z-coordinate resolution. Also, the diffusion coefficients is expected to be as small as possible to obtain a better spatial resolution.

\begin{figure}
	\centering
	\includegraphics[scale=0.25]{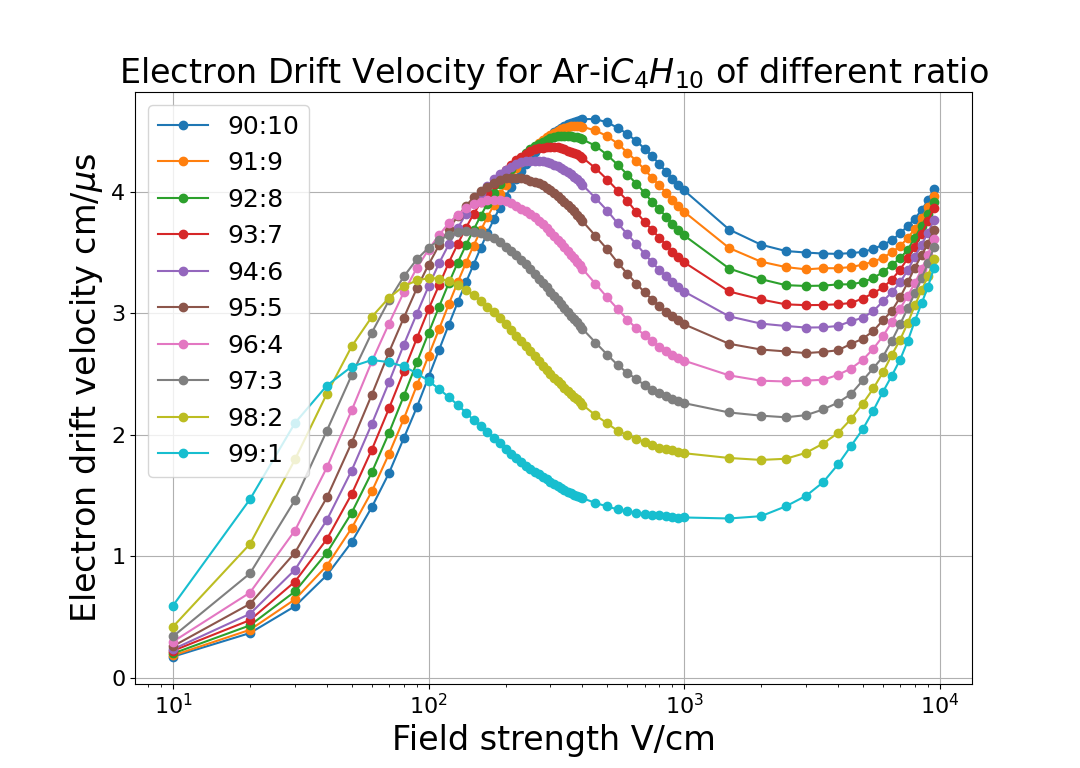}
	\includegraphics[scale=0.25]{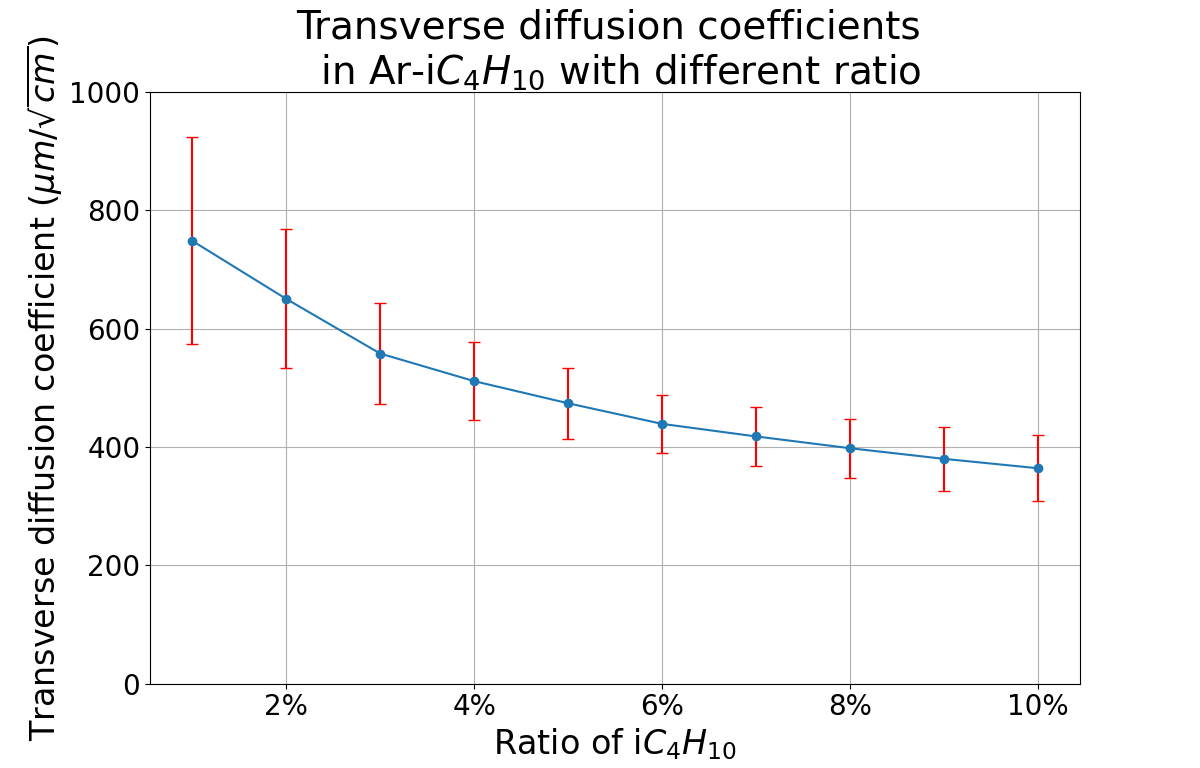}
	\caption{The drift velocity and transverse diffusion coefficients of electrons in Ar-$\rm iC_4H_{10}$ at different proportions}
	\label{Figure1}
\end{figure}

Thus, in this section, we would simulate the drift velocity and diffusion coefficients of electrons by Garfield++\cite{bib: Garfield++}. We set the temperature as the room temperature, the pressure as the standard atmospheric pressure, and the Penning coefficient as 0.57\cite{bib: Penning}. In the case of Ar-$\rm iC_4H_{10}$, we simulate the drift velocity and transverse diffusion coefficients of electrons in Ar-$\rm iC_4H_{10}$ at different proportions, shown as Fig.\,\ref{Figure1}.

The drift velocity of electrons in TPC is supposed to be stable to improve the time resolution and z-coordinate spatial resolution and to be as fast as possible to reduce the dead time of the detector, while the transverse diffusion coefficients is required to be as small as possible to improve the xy-coordinate resolution, so we choose Ar-$\rm iC_4H_{10}$ at 90:10 and constrain the electric field between 280 and 650\,V/cm where the drift velocity is between 4.4 and 4.6\,cm/$\mu$s.

\begin{figure}
	\centering
	\includegraphics[scale=0.25]{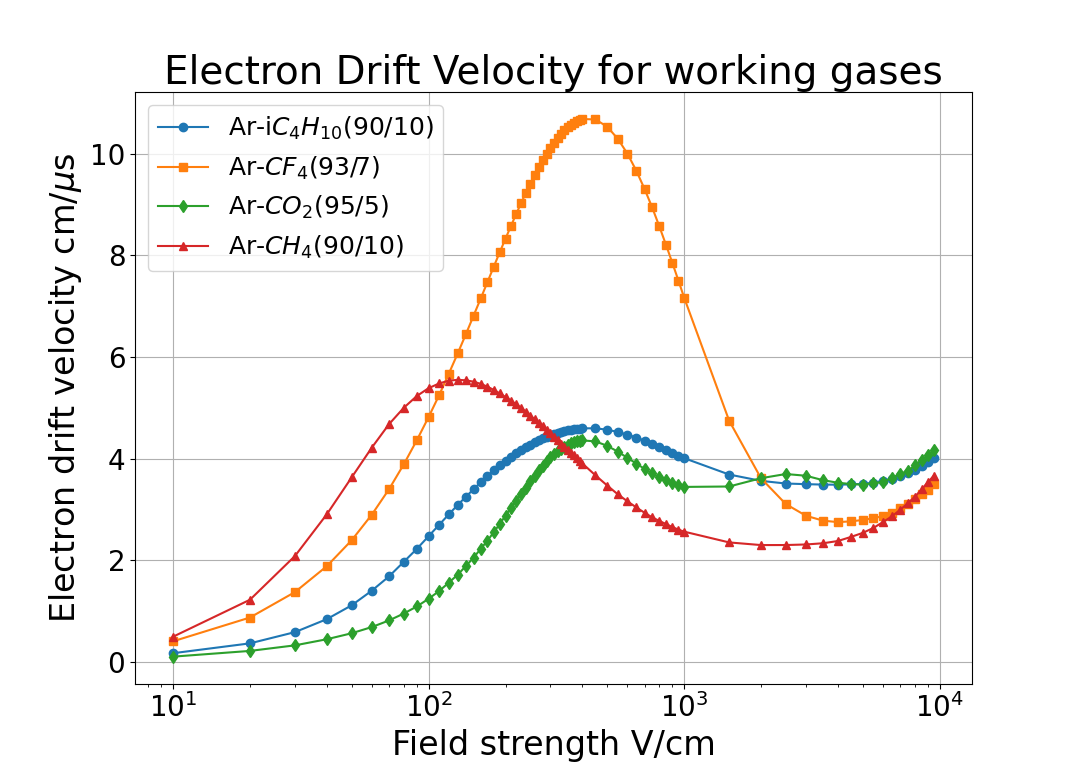}
	\caption{The drift velocity of electrons in different working gases}
	\label{Figure2}
\end{figure}

Similarly, we simulate the same properties of electrons in Ar-$\rm CF_4$, Ar-$\rm CH_4$, Ar-$\rm CO_2$ at different ratio then optimize the proportion of gases and the electric field, which is summarized as Fig.\,\ref{Figure2} and Table \ref{Table1}. The result for Ar-$\rm CH_4$ matches well with reference\cite{bib: TU-gas}. If a TPC is designed with one meter drift zone, then the corresponding drift time and xy-coordinate spatial resolution is also summarized in Table \ref{Table1}. Note that when a TPC operates under a magnetic field, the transverse diffusion coefficient is suppressed as 

\begin{equation}
	\frac{D_T(B)}{D_T(0)}=\sqrt{\frac{1}{1+\omega^2\tau^2}}
\end{equation}

where $\omega=\frac{q_e}{m_e}|B|$ and $\tau$ is the mean time between collisions.

\begin{table*}
	\centering
	\caption{The appropraite field, drift velocity and transverse diffusion coefficients in different working gases}
	\begin{tabular}{|c|c|c|c|c|c|c|}
		\hline
		& ratio & \begin{tabular}[c]{@{}l@{}} optimized \\ field / V \end{tabular} & \begin{tabular}[c]{@{}l@{}} drift velocity \\ / ($cm/\mu s$) \end{tabular} & \begin{tabular}[c]{@{}l@{}} transverse diffusion \\ coefficient / ($\mu m/\sqrt{cm}$) \end{tabular} & \begin{tabular}[c]{@{}l@{}} transverse diffusion \\ coefficient under 1 Tesla\\ / ($\mu m/\sqrt{cm}$) \end{tabular} & \begin{tabular}[c]{@{}l@{}} drift time in \\ 1 meter TPC \\ / $\mu$s \end{tabular} \\
		\hline
		Ar-$\rm iC_4H_{10}$ & 90/10 & 210$\sim$1000 & 4.0$\sim$4.6 & 340$\sim$380 & 150$\sim$300 & 21.7$\sim$25.0 \\
		\hline
		Ar-$\rm CO_2$ & 95/5 & 300$\sim$550 & 4.0$\sim$4.4 & 360$\sim$425 & 200$\sim$300& 22.7$\sim$25.0 \\
		\hline
		Ar-$\rm CF_4$ & 93/7 & 300$\sim$550 & 10.0$\sim$10.7 & 225$\sim$265 & 70$\sim$130 & 9.3$\sim$10.0 \\
		\hline 
		Ar-$\rm CH_4$ & 90/10 & 80$\sim$220 & 5.0$\sim$5.6 & 570$\sim$615 & 90$\sim$200 & 17.8$\sim$20.0 \\
		\hline
	\end{tabular}
	\label{Table1}
\end{table*}

\section{The transparency rate of electrons}

In a gaseous TPC, the energy loss of a charged particle is mainly converted to primary ionization, so the readout chamber of this TPC is expected to collect all of the primary electrons to reconstruct precise dE/dx. Thus the ratio of primary electrons passing through the multi-THGEMs, especially the top one, is required to be as large as possible.

Generally speaking, when the electrons of primary ionization reach the first THGEM, some may be absorbed by the copper of the upper and lower surfaces of THGEMs, which could not arrive lower layers of THGEMs to avalanche. So we define the effective transparent rate of electrons as the ratio at which the electrons could arrive the second THGEM.

The main factor that influence the transparency rate is the optical transparency rate that means the ratio of THGEM holes area to total THGEM working area. Since the THGEM is a regular triangle periodic structure, the optical transparency can be formulated as

\begin{equation}
	\eta_{optic}=\frac{\pi d^2}{2\sqrt{3}l^2}
\end{equation}

where $d$ is the hole diameter and $l$ is the pitch of the top THGEM.

In addition, the field lines near THGEM holes have a converging shape, so the motion of electrons near THGEM holes also converges to the holes. Then the transparent rate is generally larger than the optical transparency. Furthermore, the transparency is also influenced by electric field of both drift area and transfer area. Thus in this section, we would study the effect of the optical transparency and the electric field on the transparent rate of electrons.

In this section, we establish a model of a periodic unit of a THGEM by COMSOL. The THGEM geometry is set as the thick 0.4\,mm, the pitch 1.0\,mm and the transfer area height 2\,mm, while the THGEM hole diameters and electric field in drift and transfer zone are variables to study. The electrons of primary ionization is set as a uniform distribution on a xy-plane in the drift area.

\subsection{The influence of the ratio of hole aperture to pitch on the transparency}

\begin{figure}
	\centering
	\includegraphics[scale=0.25]{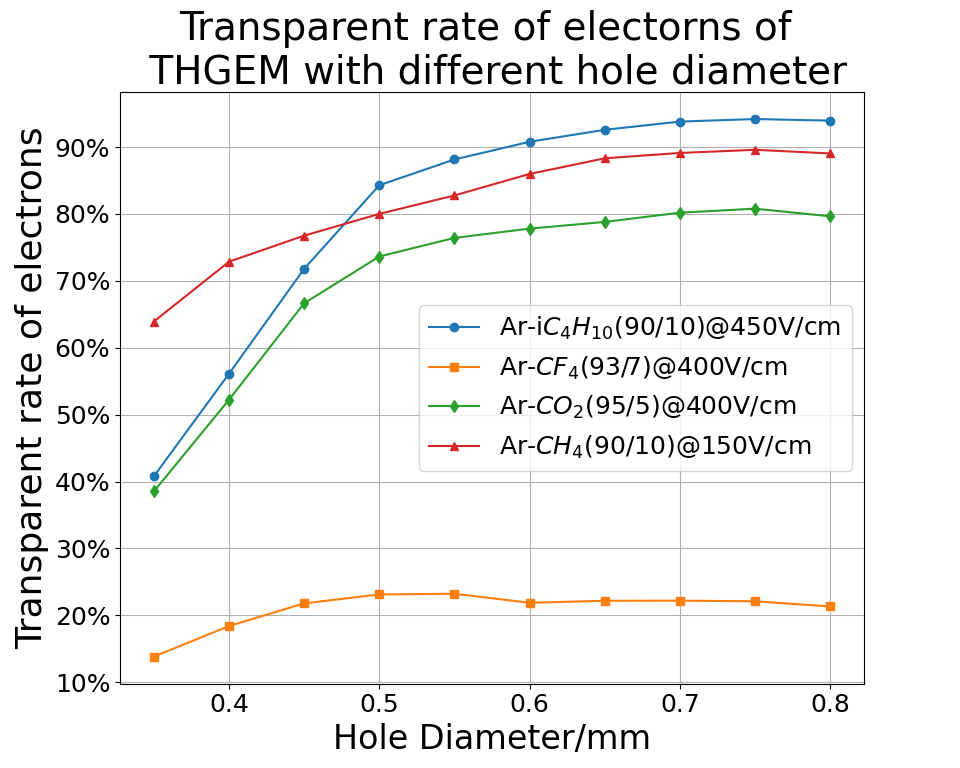}
	\caption{The Transparent rate of electrons of THGEM with different hole size}
	\label{Figure3}
\end{figure}

The electric field in the transfer area is fixed as 1500\,V/cm, and the drift field is set according to Table \ref{Table1}, while the hole diameter is a variable between 0.4 and 0.8\,mm. As for the voltage of the first THGEM, since this THGEM can only suppress the IBF of lower THGEMs, so the gain of this THGEM is supposed to be low and we require it less than 10. Finally, we set the voltage of this THGEM as 800\,V. In one periodic unit, initial electrons is set 10,000. According to the electron motion simulated by Garfield++, we count the proportion of electrons passing through the transfer area, i.e. transparency, and summarize it as Fig.\,\ref{Figure3}.

According to Fig.\,\ref{Figure3}, when the aperture is larger than 0.6\,mm, the transparency would not change significantly with the hole diameter. On the other hand, the hole size is not supposed to be too large in order to reduce the ion back flow. So it is enough to set the hole diameter as 0.6\,mm. In addition, in Fig.\,\ref{Figure3}, it should be remarked that in Ar-$\rm iC_4H_{10}$(90/10), the transparency can reach 90\%, so we will choose Ar-$\rm iC_4H_{10}$(90/10) as working gas in further simulation.

\begin{figure}
	\centering
	\subfigure[The ended points of electrons in Ar-$\rm iC_4H_{10}$]{
		\includegraphics[scale=0.1]{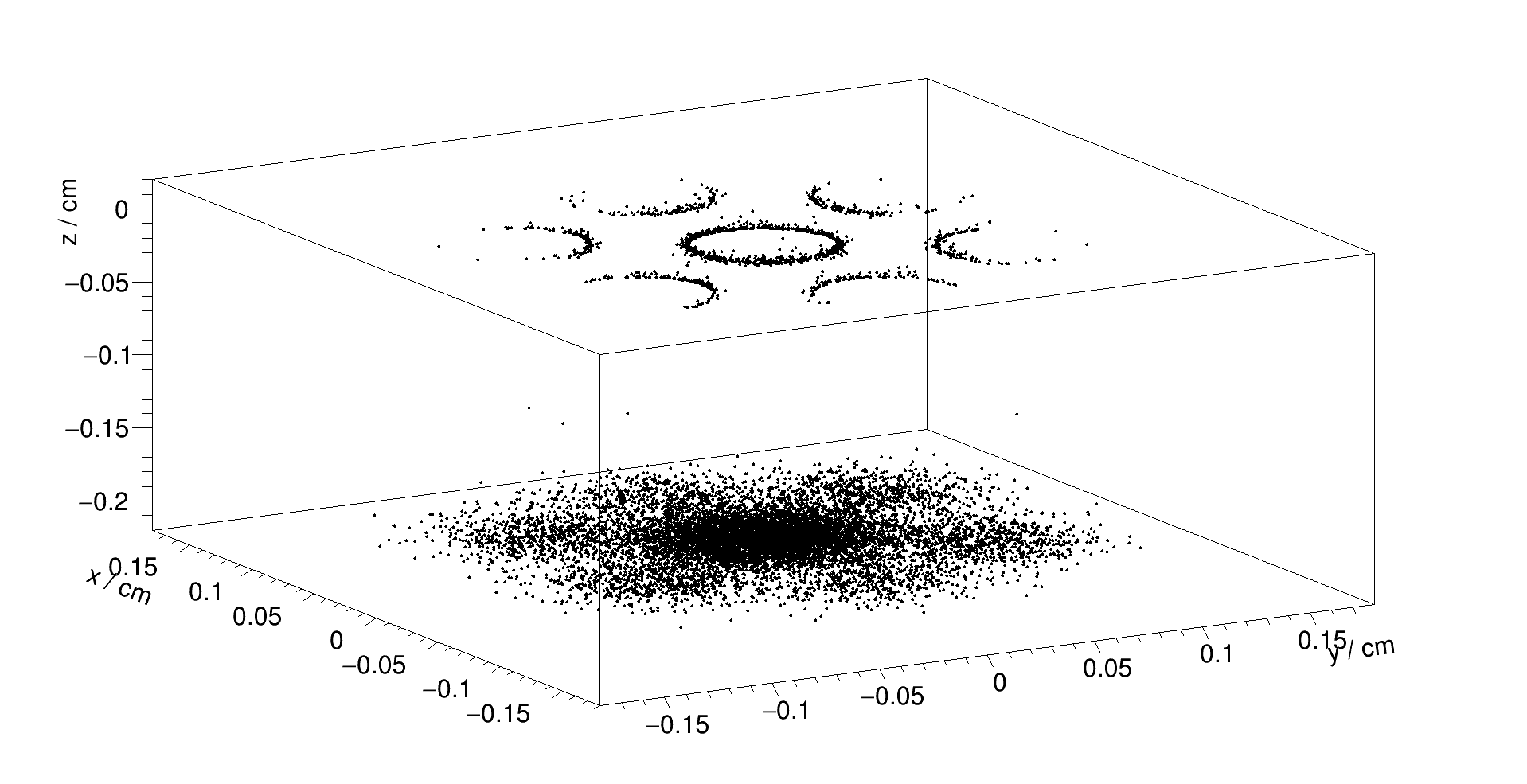}
	}
	\subfigure[The ended points of electrons in Ar-$\rm CF_4$]{
		\includegraphics[scale=0.1]{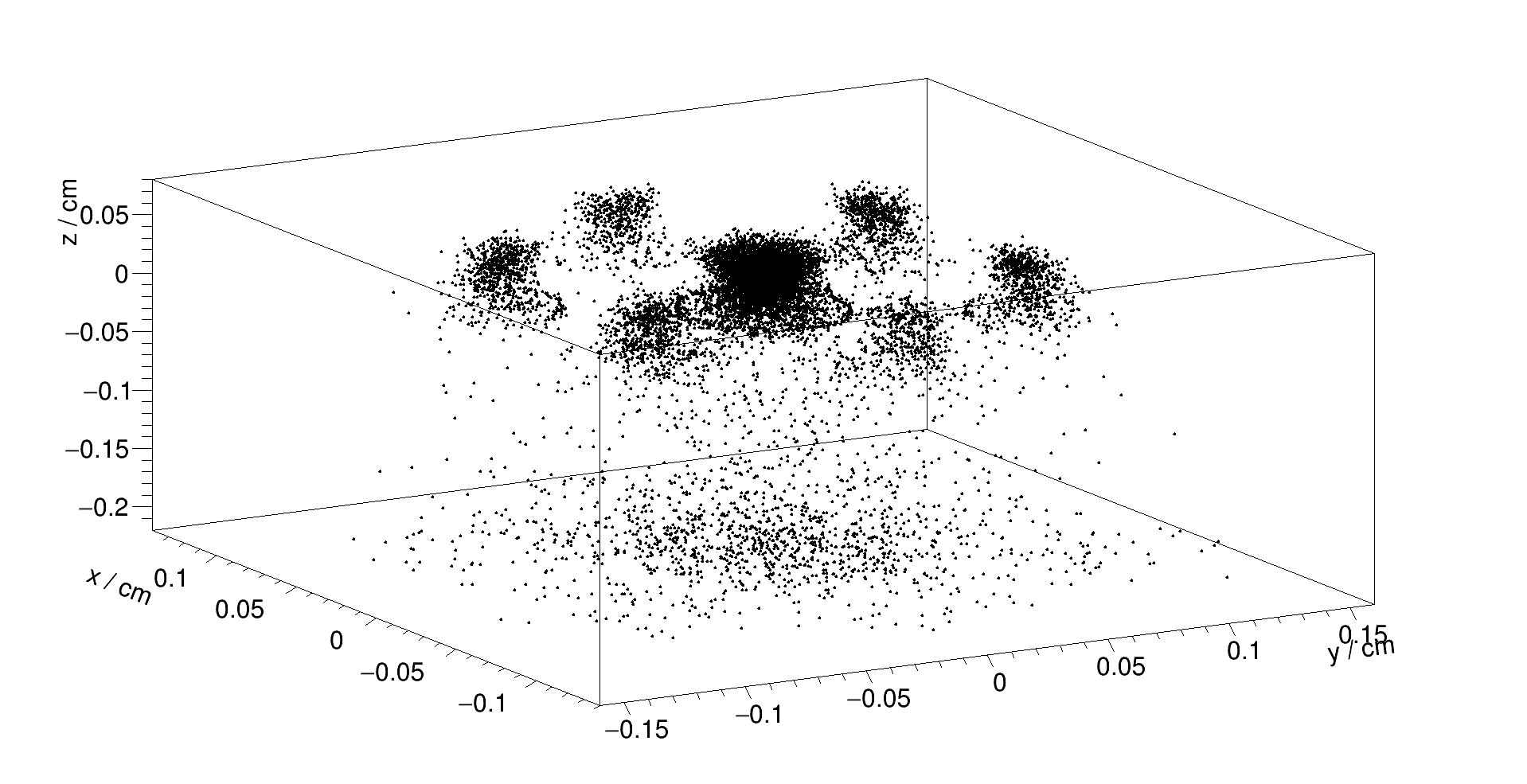}
	}
	\caption{The ended points of electrons}
	\label{Figure4}
\end{figure}

It's worth remarking that in Ar-$\rm CF_4$(93/7), the transparent rate is as low as 20\% level. By checking the end points and end status of electrons in THGEM with hole size 0.6mm in different gases, we summarize the end points in Table \ref{Table2}, and the cases of Ar-$\rm iC_4H_{10}$(90/10) and Ar-$\rm CF_4$(93/7) is visualized as Fig. \ref{Figure4}. In Ar-$\rm CF_4$(93/7), there are more than 50\% electrons that is attached in the avalanche area. The attachment effect of Ar-$\rm CF_4$(93/7) is so serious that the transparent rate is low, thus we don't choose Ar-$\rm CF_4$(93/7) as the working gas in this project.

\begin{figure}
	\centering
	\includegraphics[scale=0.25]{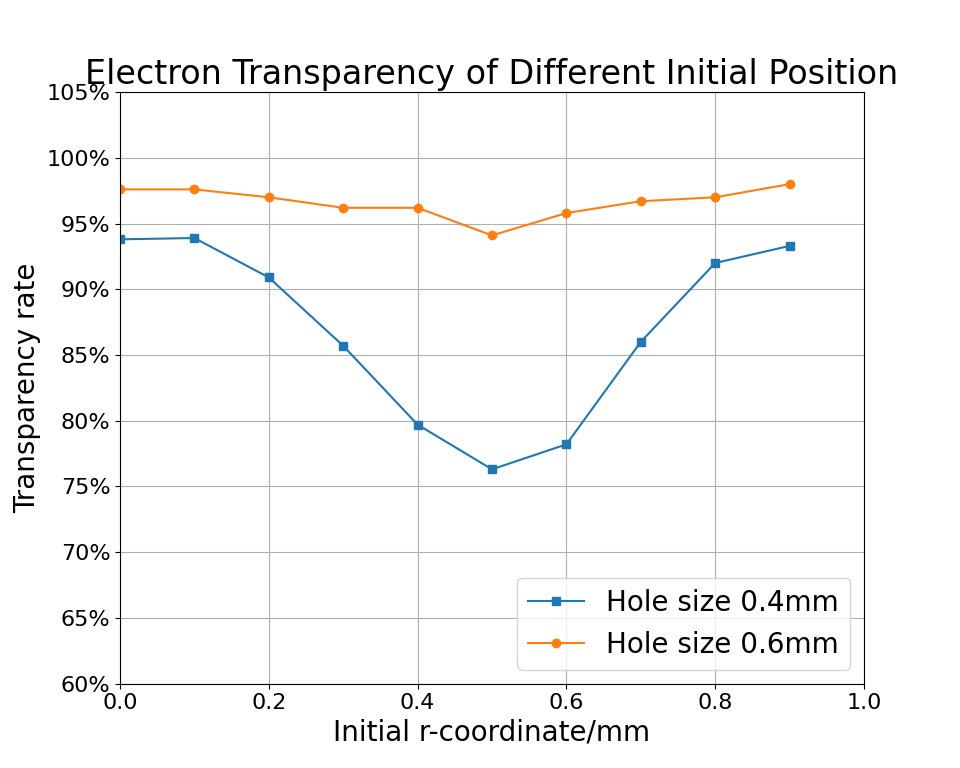}
	\caption{The transparent rate of electrons at different initial position}
	\label{Figure5}
\end{figure}

Finally, we did a further simulation on the influence of the initial position on transparency, shown as Fig. \ref{Figure5}.

\subsection{The influence of the electric field on the transparency}

Besides the optical transparency, the electric field also has a great impact on the transparent rate of electrons. The shape of electric lines near the THGEM hole depends on the ratio of field in drift and avalanche zone, and that in transfer and avalanche zone, which impacts the efficiency of electron entering and leaving holes respectively. 

\begin{figure}
	\centering
	\subfigure[Transparent rate of electrons of THGEM under different drift field]{
		\includegraphics[scale=0.25]{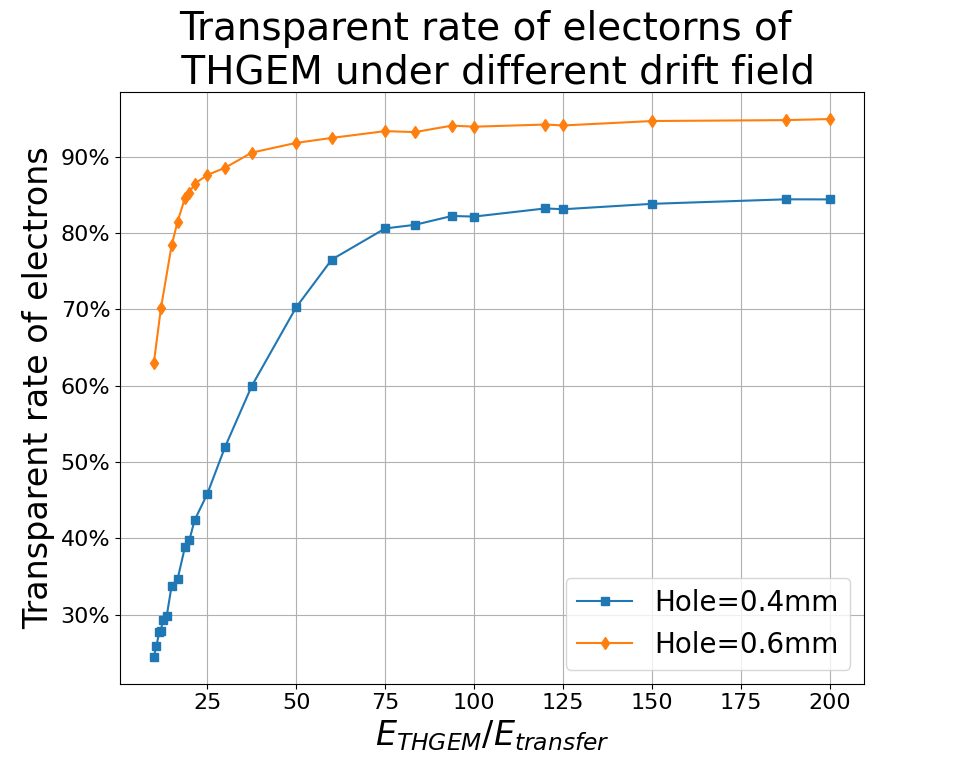}
	}
	\subfigure[Transparent rate of electrons of THGEM under different transfer field]{
		\includegraphics[scale=0.25]{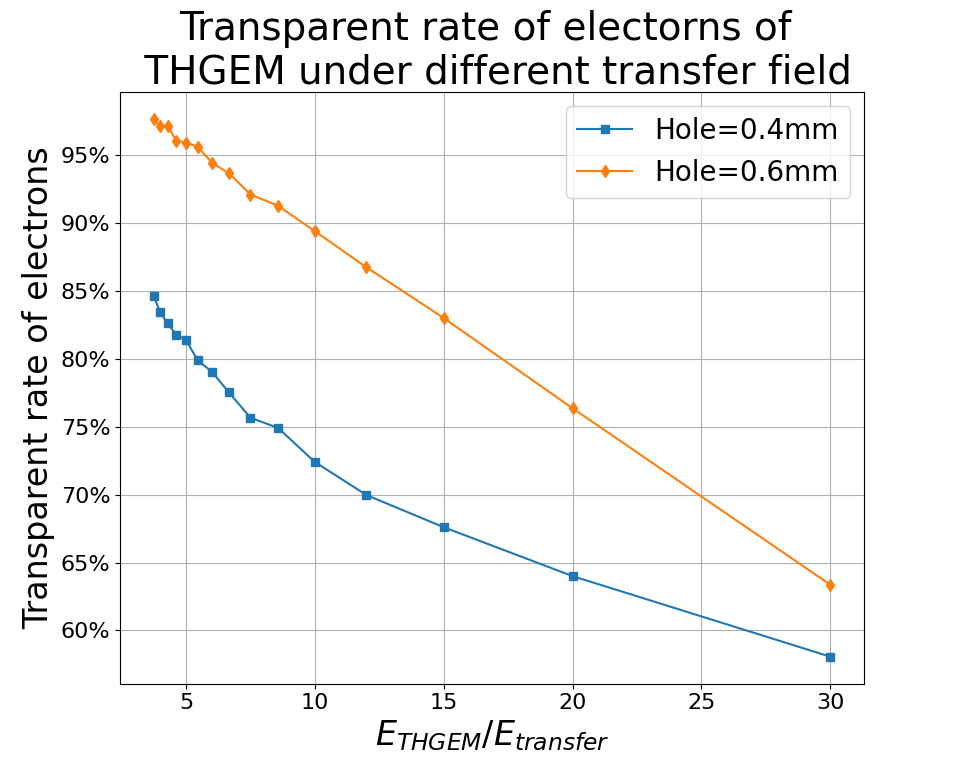}
	}
	\caption{The Transparent rate of electrons of THGEM under different drift field and transfer field}
	\label{Figure6}
\end{figure}

Since the transparent rate is generally insensitive to the exact value of the electric field strength, we may fix the voltage of THGEM to 800\,V, i.e. the avalanche field to 15\,kV/cm, and change the electric field in both drift zone and transfer zone. The result can be summarized as Fig.\,\ref{Figure6}.

According to Fig. \ref{Figure6}, to ensure the transparent rate larger than 90\%, the drift field is supposed to be less than 1/25 of the avalanche field, while the transfer field should be larger than 1/10 of the avalanche field. In the case where the avalanche field is 15\,kV/cm, the threshold of the drift and transfer field are 800\,V/cm and 1.5\,kV/cm respectively.

\begin{table*}
	\centering
	\caption{Electrons ended points}
	\begin{tabular}{|c|c|c|c|c|}
		\hline
		 & Ar-$\rm iC_4H_{10}$(90/10) & Ar-$\rm CO_2$(95/5) & Ar-$\rm CH_4$(90/10) & Ar-$\rm CF_4$(93/7) \\
		 \hline
		 upper metal & 0\% & 0.2\% & 0.05\% & 2.2\% \\
		 avalanche area & 5.25\% & 15.46\% & 9.55\% & 53.82\% \\
		 lower metal & 8.23\% & 8.54\% & 9.55\% & 3.48\% \\
		 transfer & 86.522\% & 73.53\% & 80.69\% & 21.82\% \\
		 other & 0\% & 2.26\% & 0.16\% & 18.67\% \\
		 \hline
	\end{tabular}
	\label{Table2}
\end{table*}

\section{The ion back flow of the multiple THGEM}

The ion back flow would cause the space charge effect, i.e. the distortion of the drift field. As a consequence, the drift velocity of electrons would be disturbed, and result in a worse time resolution and z-direction space resolution, while the counting rate of TPC is also limited. So the IBF rate is an essential technical indicator to research. In early gaseous detectors, Some groups studied a GEM-like structure called MHSP and its improved configurations called R-MHSP and F-R-MHSP, and suppressed the IBF as $1.5\times10^{-4}$ \cite{bib: MHSP,bib: R-MHSP,bib: F-R-MHSP}. Another GEM-like structure called Cobra is also studied to suppress the IBF \cite{bib: COBRA}. 

MicroMegas is another novel solution to suppress the IBF. The mesh of MicroMegas can naturally absorb ions, and the IBF can be easily lower 0.1\% level \cite{bib: MicroMegas-TPC}. The combination of GEM/THGEM and MicroMegas also shows a good performance on suppression of IBF \cite{bib: CEPC, bib: GEM+MicroMegas-TPC}. More recently, double-mesh MicroMegas is proposed \cite{bib: DMM-3}, and the latest IBF result is $3\sim5\times10^{-4}$ at gain $>10^6$ \cite{bib: DMM-IBF-1,bib: DMM-IBF-2}. But the large size and robustness of MicroMegas still needs further research.

Staggered multi-GEM/THGEMs \cite{bib: Staggered-THGEM-UCAS} is also a proposal on the rise. Some novel configurations are proposed recently \cite{bib: Staggered-THGEM-SDU}. However, all of the previous work only considered staggered strategies of two-layers although the researchers use triple- or quadruple-THGEM structures. Therefore, in this section, we would propose several staggered structures of multi-THGEMs, and study the IBF of these configurations. 

\begin{figure}
	\centering
	\subfigure["ABA" configuration]{
		\includegraphics[scale=0.1]{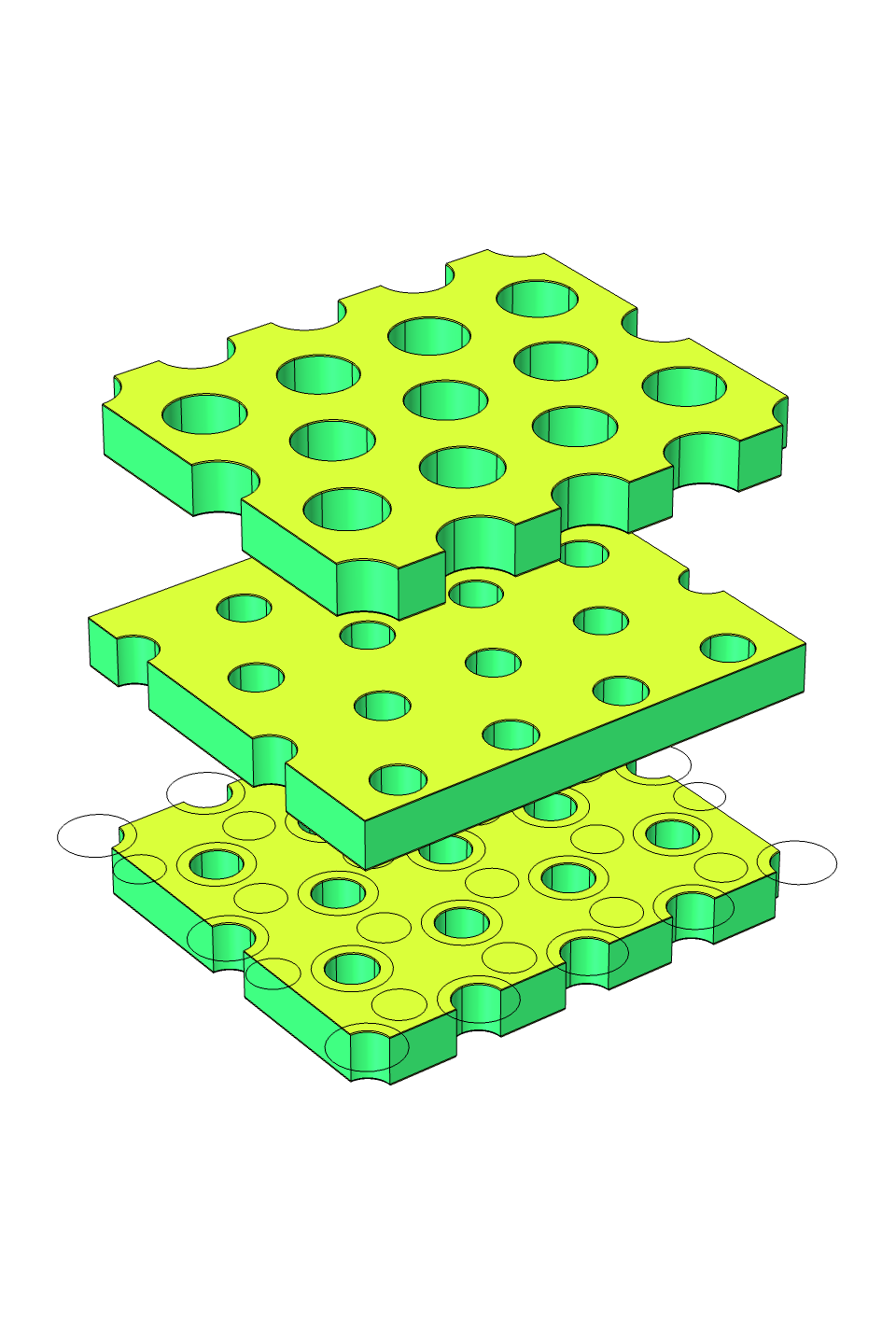}
	}
	\subfigure["CBA" configuration]{
		\includegraphics[scale=0.1]{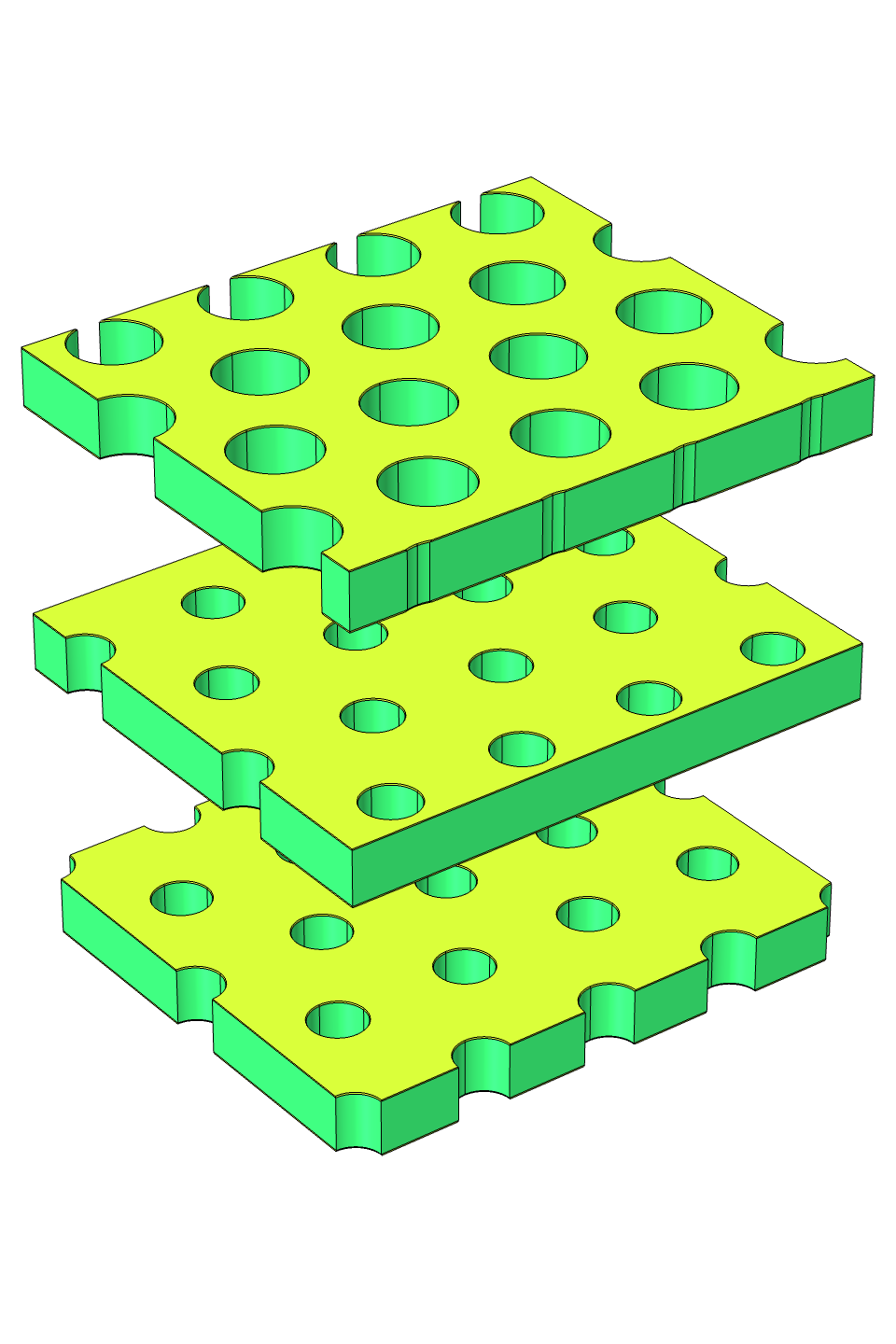}
	}
	\subfigure["BABA" configuration]{
		\includegraphics[scale=0.1]{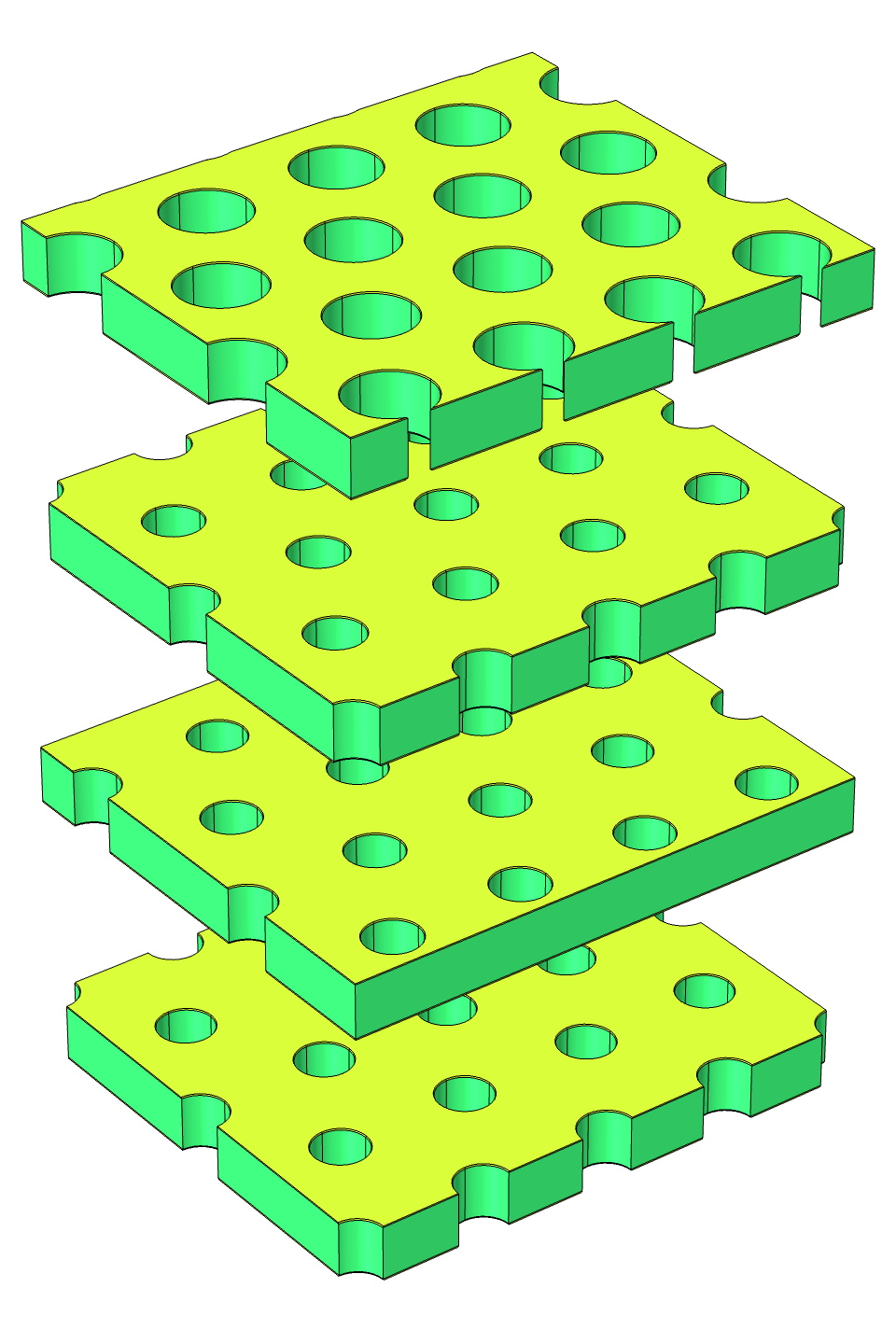}
	}
	\subfigure["CABA" configuration]{
		\includegraphics[scale=0.1]{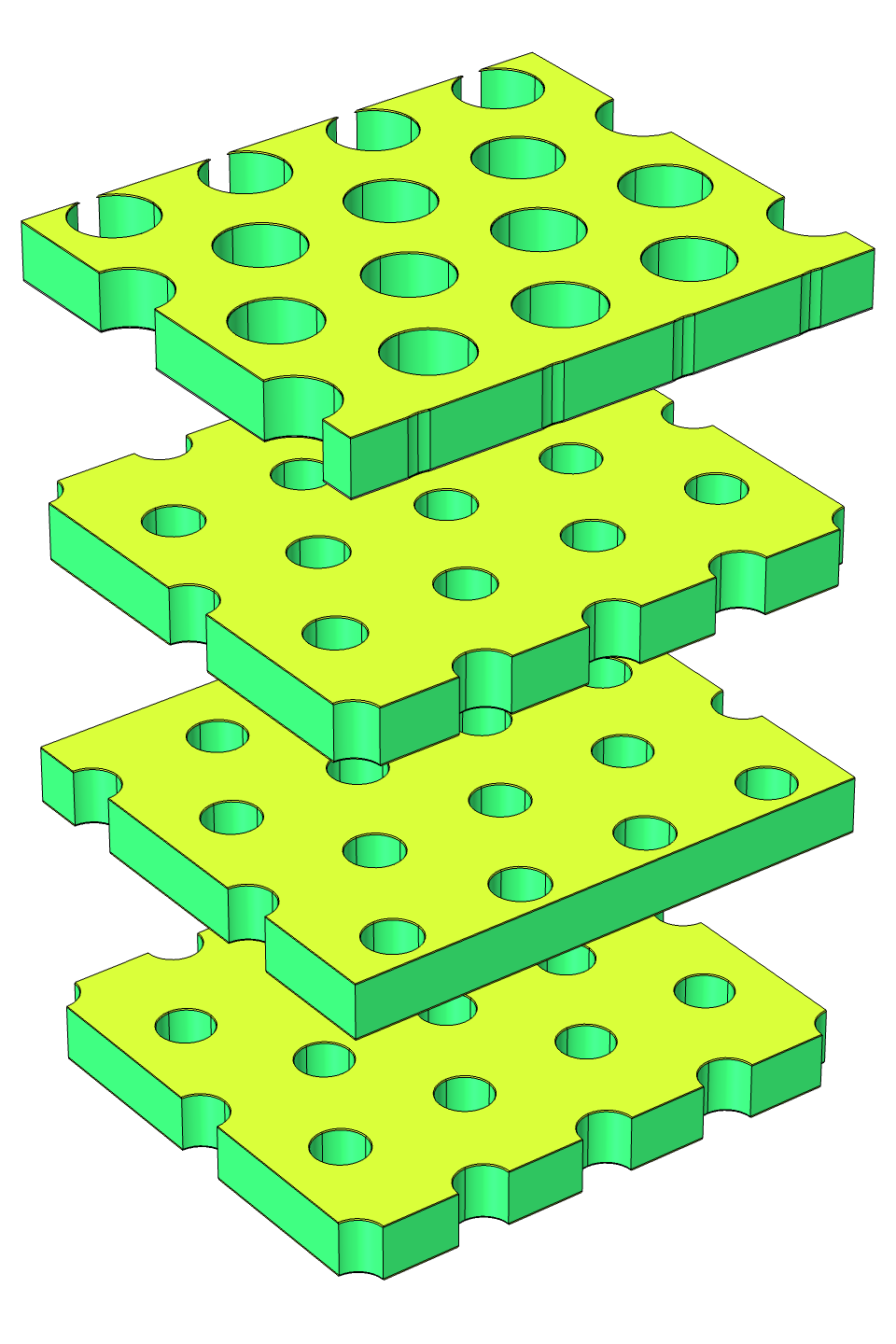}
	}
	\subfigure["ACBA" configuration]{
		\includegraphics[scale=0.1]{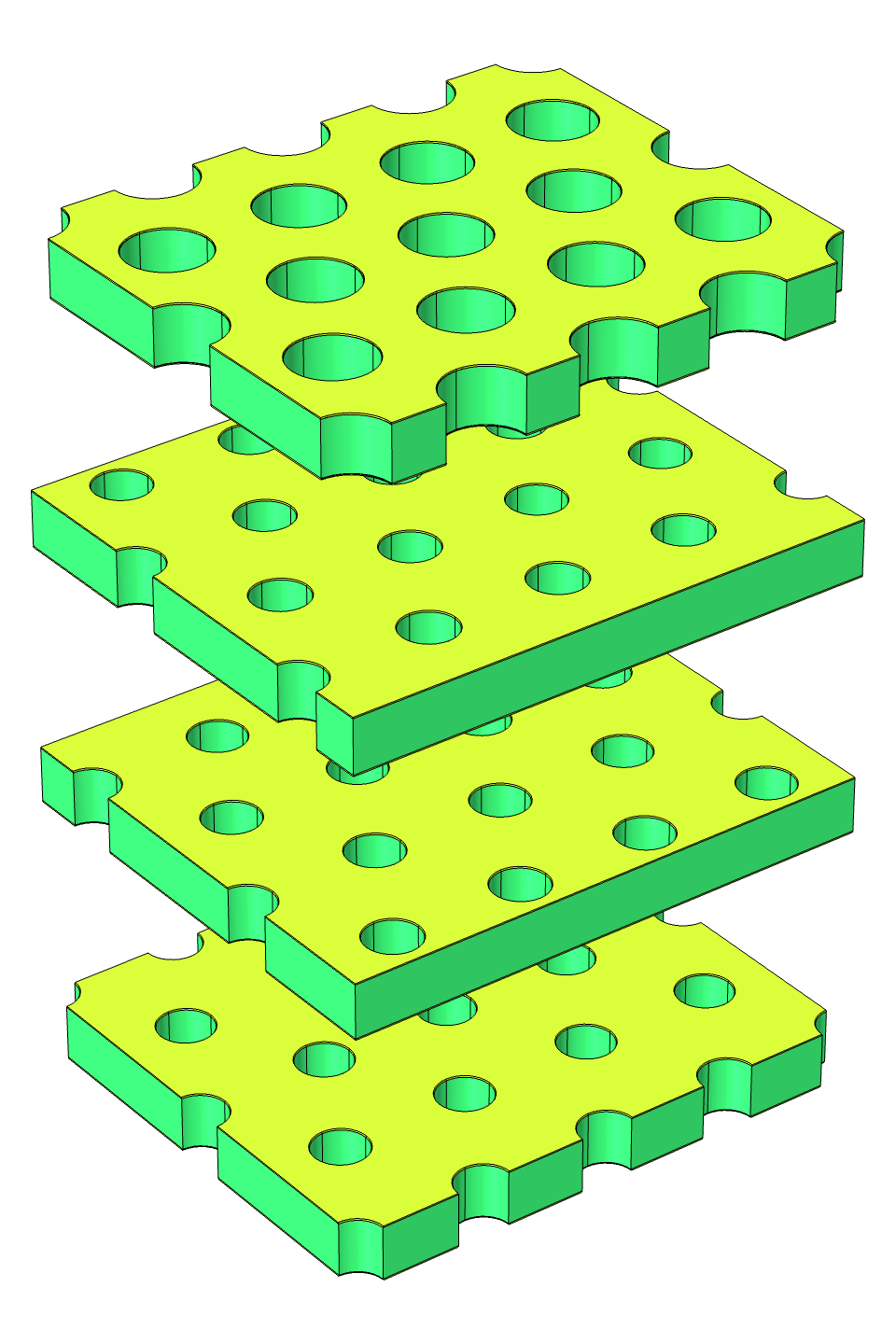}
	}
	\caption{The upper 2 figures are 2 typical configurations of triple THGEM denoted as 'ABA' and 'CBA' structures, and the lower 3 figures are 3 typical configurations of quadruple THGEM denoted as 'BABA', 'CABA' and 'ACBA' structures, where 'A' always represents the configuration of the bottom THGEM. In all figures, the yellow and blue circles on the top surface of the bottom THGEM is the projection of holes in THGEM of upper layers.}
	\label{Figure7}
\end{figure}

The fundamental staggered THGEM for 2-layers is easy to achieve by "flip flop" or 180 rotation. Then for the triple and quadruple THGEMs, the staggered strategies is the combination of these two structures. There are 2 configurations for triple-THGEM and 3 configurations for quadruple-THGEM, shown as Fig.\,\ref{Figure7} together with their periodic unit. Following the notation in solid state physics, we use symbols like "ABA" to represent a certain configuration. To avoid duplication of tags, we use a convention that the bottom two THGEMs are always marked as "A" and "B". The size of THGEM is chosen: the thickness is 0.4\,mm and the pitch is 1.0\,mm, while the hole size of top THGEM and other THGEM are 0.6\,mm and 0.4\,mm.

\begin{figure}
	\centering
	\subfigure[z-coordinate]{
		\includegraphics[scale=0.25]{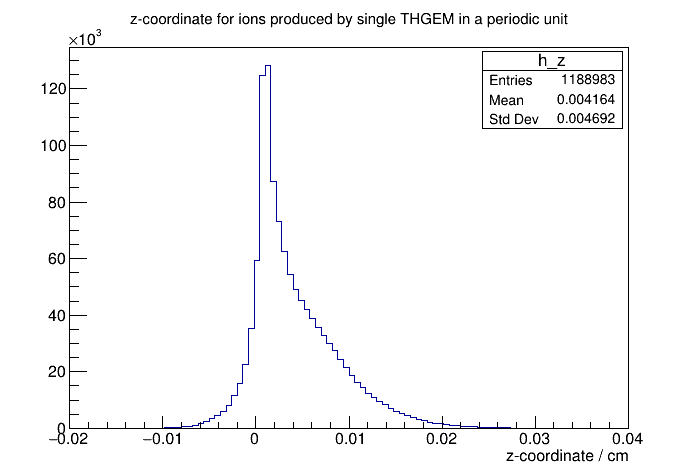}
	}
	\subfigure[r-coordinate]{
		\includegraphics[scale=0.25]{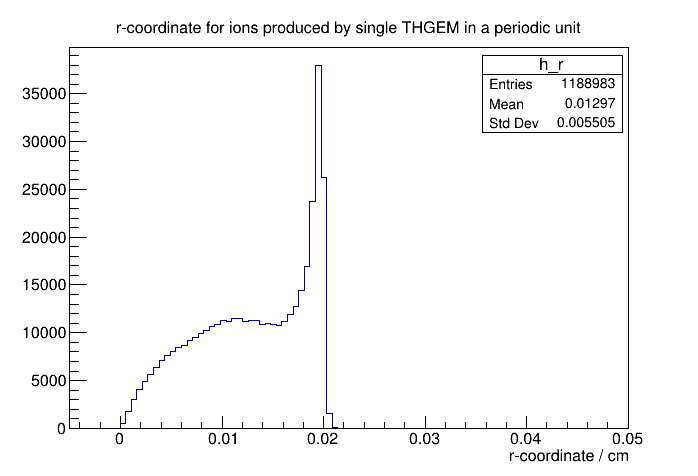}
	}
	\caption{The coordinates distribution of ions production for a single THGEM, where $z=0$ is defined as the bottom surface of THGEM, and $r=0$ means the center of a hole of THGEM}
	\label{Figure8}
\end{figure}

\begin{figure}
	\centering
	\includegraphics[scale=0.25]{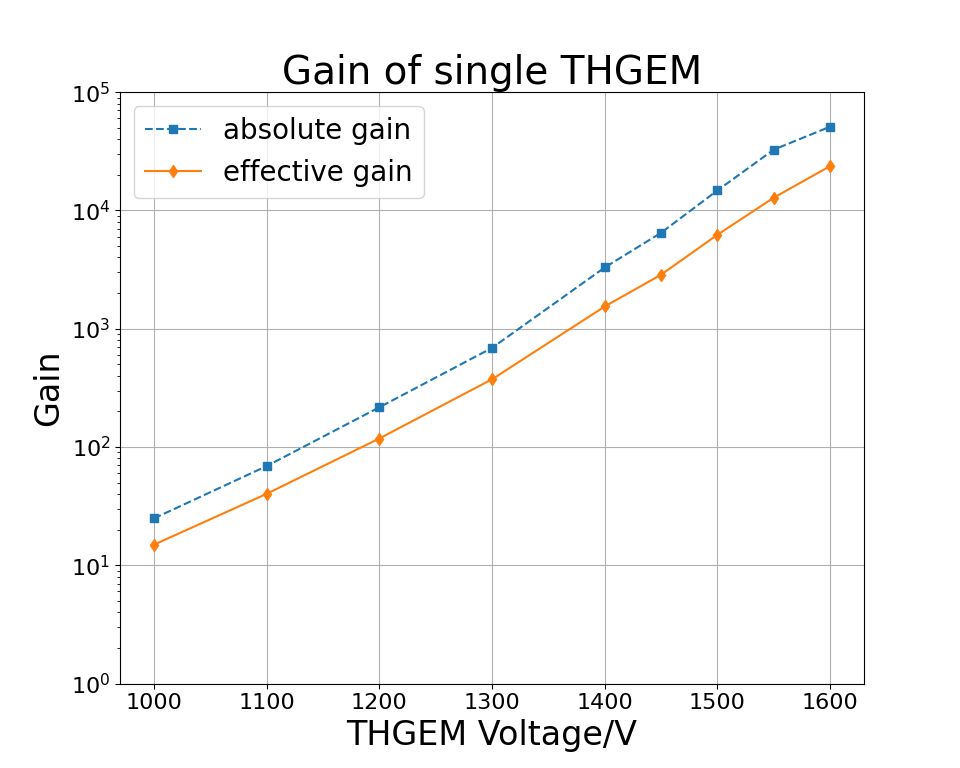}
	\caption{The gain of single THGEM}
	\label{Figure9}
\end{figure}

In order to perform Monte Carlo simulation of ion back flow, we need to know the position distribution of ions produced at a THGEM. It’s feasible to assume that the production coordinates are uniform distribution in the holes of a single THGEM, although this algorithm doesn't contain any information of avalanche procedures. Thus we simulate the avalanche process in one hole of THGEM, and record the production position distribution of ions, including the z-direction and r-direction distribution, shown as Fig.\,\ref{Figure8}. Besides, the gain of single THGEM can be also obtained, shown as Fig.\,\ref{Figure9}, where the absolute gain is defined as the number of the secondary electrons in the avalanche zone, while the effective gain is defined as the number of the electrons that can achieve the anode and induce a electronic signal.

In order to suppress the IBF by using the upper layers of THGEM, the gain of the upper layers of THGEM should not be too large so that the ions are mainly produced in the bottom THGEM. So the ions produced at THGEM of upper layers are negligible comparing to that at bottom one, and it's feasible only simulate the ions produced at the bottom THGEM. 

\begin{figure}
	\centering
	\includegraphics[scale=0.25]{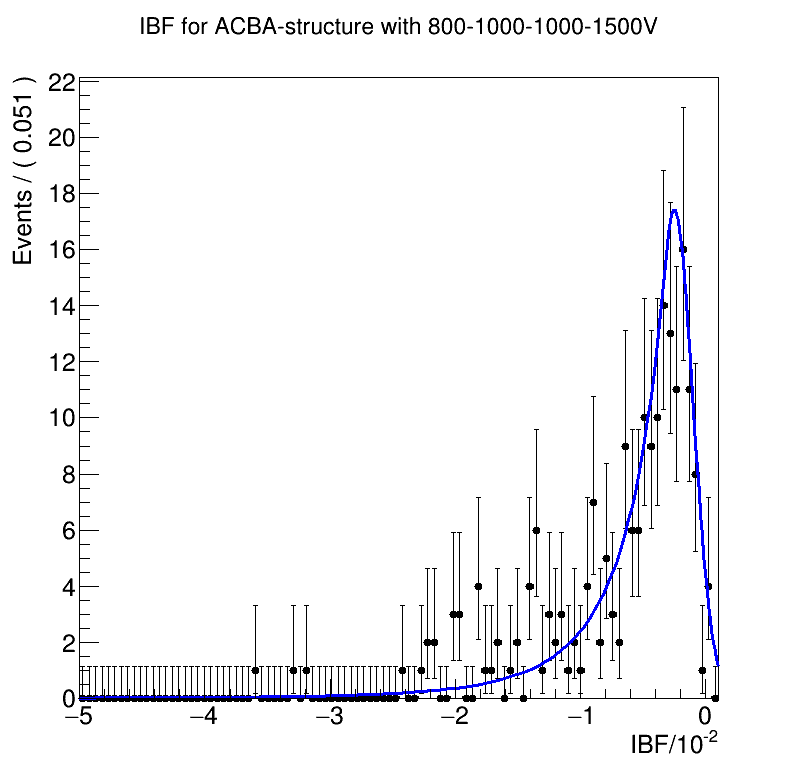}
	\caption{The IBF histogram of "ACBA" structure with 800-1000-1000-1500V}
	\label{Figure10}
\end{figure}

Commonly-used electronics can generally read out the charge of the $0.1\sim10$\,pC magnitude, so the total gain of the multi-THGEM should be $10^5\sim10^7$ level. On the other hand, the total gain is expected to mainly come from the contribution of the bottom THGEM so that the upper THGEMs are able to absorb more ions. Therefore, for triple THGEM structures, we set the gains of each THGEM are $10^0$, $10^2$ and $10^4$, while for quadruple THGEM structures, we would like to set the gains of each THGEM are $10^0$, $10^1$, $10^1$ and $10^4$. According to Fig. \ref{Figure9}, the corresponding voltage should be 800\,V, 1200\,V and 1500\,V for triple THGEM, while 800\,V, 1000\,V, 1000\,V and 1500\,V for quadruple THGEM. The hole diameter of top THGEM is set as 0.6\,mm and other THGEMs are 0.4\,mm. The field of both transfer and induction zone are set as 2\,kV/cm. For a single event, the IBF simulation results are summarized as Table \ref{Table3} for triple THGEM and Table \ref{Table4} for quadruple THGEM. The IBF of more events can be filled in histograms and we use crystal ball function to fit, an example of "ACBA" structure is shown as Fig. \ref{Figure10}.

According to the Table \ref{Table3}, we can conclude that the ions absorbed by the lower two THGEM are similar in both configuration of triple THGEM. The difference is that the top THGEM of "CBA" configuration can absorb almost half ions remained of the lower two THGEM, while that of ABA configuration can absorb almost none. But the IBF of both configuration is larger 5\% level. So we have to consider quadruple THGEM.

According to the Table \ref{Table4}, the IBF of configuration "ACBA" can be reduced to 0.5\%, which is similar to double MicroMegas structure\cite{bib: DMM-IBF-1, bib: DMM-IBF-2}. As for the other two configurations, the IBF of both configurations are 3\% level, which are too large comparing with the ACBA configuration.

\begin{figure}
	\centering
	\includegraphics[scale=0.25]{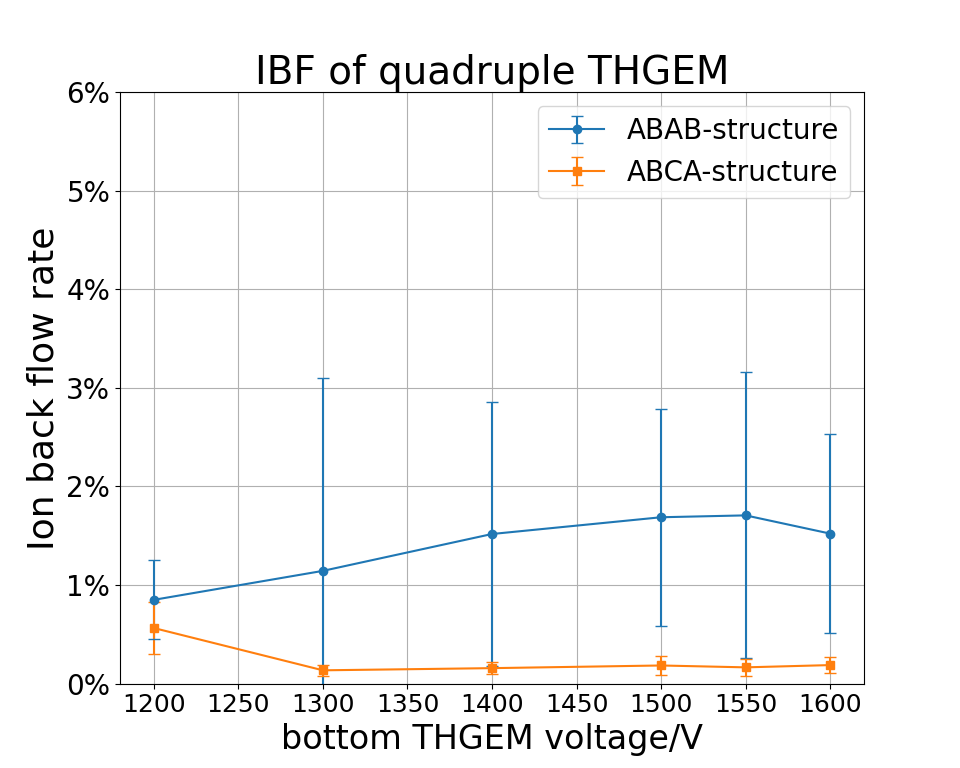}
	\caption{The IBF ratio under different voltages of the bottom THGEM}
	\label{Figure11}
\end{figure}

Finally, we study the influence of voltage of the bottom THGEM on the IBF. The voltage of upper layers of THGEMs as 800\,V, 1000\,V and 1000\,V, and the simulation result is summarized as the Fig.\,\ref{Figure11}. The IBF can be reduced as low as 0.2\% in the "ACBA" configuration while the IBF in traditional staggered strategy "BABA" is 2\% level. This means the "ACBA" structure can reduce the IBF by a factor of $10^{-1}$.

\begin{table}
	\centering
	\caption{The Ions of each THGEM in triple-THGEM structures}
	\begin{tabular}{|c|c|c|}
		\hline
		Configuration & ABA & CBA \\
		\hline
		IBF & 8.15\% & 4.17\% \\
		THGEMtop & 0\% & 3.98\% \\
		THGEMmid & 44.05\% & 44.15\% \\
		THGEMbottom & 47.80\% & 47.70\% \\
		\hline
	\end{tabular}
	\label{Table3}
	
	\caption{The Ions of each THGEM in quadruple-THGEM structures}
	\begin{tabular}{|c|c|c|c|}
		\hline
		Configuration & BABA & CABA & ACBA \\
		\hline
		IBF & 3.24\% & 3.31\% & 0.46\% \\
		THGEM1 & 5.82\% & 5.03\% & 1.85\% \\
		THGEM2 & 4.69\% & 4.38\% & 10.41\% \\
		THGEM3 & 27.49\% & 27.44\% & 27.47\% \\
		THGEM4 & 59.87\% & 59.84\% & 59.83\% \\
		\hline
	\end{tabular}
	\label{Table4}
\end{table}

\section{Conclusion}

In this paper, we propose to use large-hole-diameter THGEM to improve the transparency of electrons and to use staggered multi-THGEM structure to suppress the IBF. The simulation by Garfield++ shows that the transparency rate can reach more than 90\% in Ar-$\rm iC_4H_{10}$ and the IBF can be suppressed to 0.2\% level in "ACBA" configuration of quadruple THGEM. 

It's remarkable that our algorithm to simulate the IBF is that firstly study the avalanche procedure in single THGEM to obtain the probability density function(PDF) of the ion production position, then drop ions based on this PDF. Comparing to uniform distribution of initial ions, this algorithm contains more information of the avalanche procedure, although it needs more powerful computing source. Further, if the computing power is even better, one can attempt to do a whole simulation of multi-THGEMs, i.e. one can directly study the avalanche procedure of the primary electrons and the drift procedure of ions in multi-THGEM structures. 

In a large-scale experiment based on TPC, beyond indicators like electron transparency and IBF, other engineering issues including mechanical stability, service life and large-size manufacturing should also be considered, and these are the advantages of THGEM. Further, in recent years, there are some novel deformations of THGEM such as THCOBRA\cite{bib: THCOBRA} and M-THGEM\cite{bib: M-THGEM} which can suppress the IBF effectively. Therefore, it is worth attempting to use THGEM, THCOBRA and M-THGEM with multi-layer strategies in future TPC.

\section*{Acknowledgement}

This work was supported by the National Natural Science Foundation of China (Grant Nos. 11575193, U1732266, U1731239, 12027803), Key Research Program of Frontier Sciences, CAS, Grant No. QYZDB-SSW-SLH039.






\end{document}